# Complexity Control

**Korosh Mahmoodi**[1,*], **Scott E. Kerick**[1], **Piotr J. Franaszczuk**[1,2], **Paolo Grigolini**[3], **and Bruce J. West**[3,4]

[1]US Combat Capabilities Development Command, Army Research Laboratory, Aberdeen Proving Ground, MD 21005, USA
[2]Department of Neurology, Johns Hopkins University School of Medicine, Baltimore, MD 21287, USA
[3]Center for Nonlinear Science, University of North Texas, Denton, TX 76203, USA
[4]Office of Research and Innovation, North Carolina State University, Raleigh, NC 27695, USA
[*]koroshmahmoodi@gmail.com

## ABSTRACT

We introduce a dynamic model for complexity control (CC) between systems, represented by time series characterized by different temporal complexity measures, as indicated by their respective inverse power law (IPL) indices. Given the apparent straight forward character of the model and the generality of the result, we formulate a hypothesis based on the closeness of the scaling measures of the model to the empirical complexity measures of the human brain. The main differences between the empirical Complexity Matching and Management Effect (CMME) model and the current CC model are: 1) CC is based on reinforcement learning (RL), while CMME is a quantum mechanical effect based on linear response theory. 2) CC depends on the interaction's nature, while CMME is independent of the strength of the perturbation 3) In CC, the complexity of the interacting networks changes over time whereas in CMMEs, they remain unchanged during the perturbation. 4) CC is manifest on single organ-network (ON) time series (ONTS), while CMMEs only appear at the level of large ensemble averages and at the asymptotic regime. Consequently, CC is a proper model for describing the recent experimental results, such as the rehabilitation in walking arm-in-arm and the complexity synchronization (CS) effect between the ONTS. The CC effect can lead to the design of mutual-adaptive signals to restore the misaligned complexity of maladjusted ONs or, on the other hand, to disrupt the complexity of a malicious system and lower its intelligent behavior.

## 1. Introduction

### 1.1 Requisite Variety

In the decade following the end of World War II, scientists that had during the war been devoting their talents to the encryption and decryption of codes, enhancing signal-to-noise ratios in various sensors, and increasing the lethality of weaponry, turned their attention to the metaphoric beating of their intellectual swords into ploughshares. The mathematician Norbert Wiener created a new branch of science he called Cybernetics[1], to study the interaction between humans and machines; his young friend and colleague Claude Shannon put the fledgling science of communication on a firm mathematical foundation with the introduction of the entropy definition of information[2] as had Wiener in cybernetics; another friend John von Neumann with Oskar Morgenstern conceived and brought to fruition Game Theory, as a novel way to quantify social interactions[3]. These three perspectives were synthesized by W. Ross Ashby in his discussion of how to control and regulate complex systems leading to what he called "The Law of Requisite Variety", which we refer to herein as Ashby's Law[4]:

> Any system that governs another, larger complex system must have a degree of complexity comparable to the system it is governing.

As an example of the interaction of two complex networks, consider how we respond to a speaker delivering a lecture. An engaging speaker does not stand in one spot and talk in a monotone, but walks about the stage, arms moving to emphasize the words, with the timber and amplitude of their voice rising and falling to punctuate what is being said. What is said at any given moment is only part of the information being transmitted and is not necessarily the most impactful part. The way a person is dressed, how their body moves, their facial expressions, as well as their voice and choice of words, are all different information modalities, separate and distinct from the cognitive content of the speech. If we like a person's appearance and how they present themselves, we are generally more likely to respond favorably to what they say.

On the other hand, if they begin by insulting our sex, race, ethnicity, or any fundamentally held belief, we may not even hear the rest of what they say, even though we stay in the lecture hall. The time a speaker spends preparing the content of the talk may have less to do with how much information we absorb than does their presentation, because as in most interactions in life: "It's not what you say, it's the way that you say it.".

This is an example of Ashby's Law. The complexity of the presentation, including content, in the above exemplar, must at least match the complexity of the individual, who is aware of the surrounding audience's reaction to what is being said, if they are to maximize the information transfer from the speaker to the audience member.

Unfortunately, this surprising law has not managed to capture the attention of the broader scientific community. It is unfortunate because the recognition of its significance was not fully appreciated until it was rediscovered nearly a half century later[5] while studying the exchange of information among dynamic complex networks[6] as the Complexity Matching and Management Effect (CMME). Here, a complex network can be anything from an organization to an organism, including one of the most complex networks, that of a human brain.

**1.2 Temporal Complexity**

One measure of the complexity of a network is the index of the inverse power law (IPL) probability density function (PDF). Such IPL PDF are ubiquitous, for example, they appear in the rank ordering of words given by Zipf Law[7], in the distribution of city sizes within a country given by Auerbach's Law[8], in the distribution of wealth captured by Pareto's Law[9] and 47 other examples across 10 disciplines listed by West and Grigolini[10].

We argue that complexity can be expressed in terms of crucial events (CEs), which are a subset of renewal events (REs) introduced into statistical analysis by Feller[11] and have identically distributed independent (*IDI*) random events. The CE subset was subsequently identified as those REs that have IPL distributions generated by the processes of spontaneous self-organized temporal criticality (SOTC)[12,13].

The complexity of CEs is defined by the waiting-time of the intervals between consecutive CEs, which have an IPL PDF $\psi(\tau)$ proportional to $\tau^{-\mu}$ with the IPL index $\mu$ in the interval (1,3). This temporal complexity has been shown to exist in physical and physiological systems such as earthquakes[14], heartbeat[15], and brain signals[16].

In[17,18] we have shown that the MFD is equal to the scaling index $D(t) = \mu(t)$, thereby enabling the direct transfer of information between ONs to be achieved using computational models. This notion of CMME evolved to include the influence of one ON on another when the level of complexity of the two networks is very different, which brings us to the recently developed concept of complexity synchronization (CS)[17,18].

**1.3 Complexity Matching and Management Effect (CMME)**

West et al.[19] identify the complexity of a network's time series with its multifractality as determined by its time-dependent multifractal dimension (MFD) $D(t)$, which enables us to determine in-advance the roles played by each network during an information exchange; that is to say, the organ-network (ON) with the greater instantaneous MFD ($D_>(t)$) will transmit the information and the ON with the lesser instantaneous MFD ($D_<(t)$) will receive the information independently of all other considerations.

Figure 1 is the set of asymptotic cross-correlations evaluated by ensemble averaging of two weakly interacting CE time series, the CE driver $X_D(t)$ and the CE driven $X_S(t)$,[20]. This cross-correlation-cube (CCC) is an encapsulation of how the two CE time series exchange information depending on the relative sizes of their complexity measures, these being the IPL indices $\mu_D$ and $\mu_S$, respectively. Both the $X_D(t)$ and $X_S(t)$ time series consist of sequences of CE IPL of randomly chosen +1's and -1's and therefore the information transfer is driven solely by the difference in the complexity of the two systems, that is, by their relative statistics.

Note that the CCC denotes the analytical solution of the cross-correlation function at the asymptotic limit and is also evaluated using ensemble averages over many trajectories. So, while CCC can't explain the strong interaction between ONs, such as brain and heart, where the statistics of the systems (classifiable by their complexity index $\mu$) change over time, but displays several remarkable properties, some of which only reach their full significance outside the realm of medical applications, see West and Grigolini (WG)[21] to explore some of the non-medical implications of the CCC.

**1.4 Complexity Synchronization (CS)**

Synchronization is today identified as the mechanism needed to coordinate activities among events in any complex, multi-level, multi-element ON. However, the mathematical concept of synchronization changes as the dynamics of the interacting ONs become increasingly more complex, which is demonstrably true of the network-of-ONs (NoONs) where the NoON structure constitutes the infrastructure of the human body. The changes in synchronization facilitate the need to coordinate activities across increasingly different time scales, from the microscopic time scales of the neural ONs within the brain, to the mesoscopic time scales of the cardiac and respiratory ONs, to the macroscopic time of circadian rhythms.

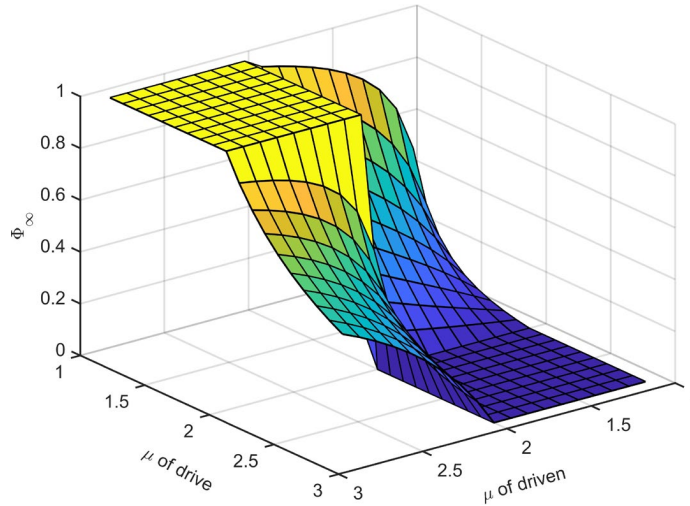

**Figure 1.** Complexity Matching and Management Effect (CMME). The z axis is the asymptotic limit of the Cross-Correlation-Cube (CCC) $\Phi(t = \infty)$ for different values of $\mu \in (1, 3)$ parameter of the driver $\mu_D$ and driven $\mu_S$ systems using the analytical solution.

On top of the ordinary synchronization described above, it has recently been shown that the temporal complexity measure $\mu$ of time series $X(t)$ data from EEG, ECG, and Respiratory ONs are time-dependent, rather than being fixed, and are in synchrony with one another[17,18], where the $\mu$ values were connected to the $\delta$ values (evaluated using the Modified Diffusion Entropy Analysis, MDEA) by $\mu = 1 + 1/\delta$. This time dependence is a consequence of the fact that the measure of complexity index $\mu$ is a measure of the MFD since we identified this IPL index with the fractal dimension in Section 1.1.

The time-dependent change in the fractal scaling of time series indicates the change in complexity of the ONs as various physiological functions are performed with the more complex ON transferring information to the less complex ON, but these transmitter and receiver roles change with the functions being performed, as well as in time. For example, information is readily exchanged among overlapping memory areas of the human brain, which are temporally heterogeneous. At any point in time, one region of the brain can receive information from sensor ONs, process that information, and transmit the resulting signal to the appropriate physiological ON for action, depending on their function and instantaneous relative complexities[22].

### 1.5 Complexity Control

To understand the dynamics of the true complexity of ONs that interact strongly (rather than with weak perturbations, as is the case in CMME) with each other, causing their complexity to change over time, we must be able to identify the most significant aspects of their empirical information exchange. For example, answering such questions as: What is the dominant or determinate factor in the complexity synchronization (CS) effect or in the restoration of complexity in walking arm-in-arm? To answer these types of questions, we need to go beyond the limits of the CMMEs, i.e., beyond the linear response theory, ensemble averages, and asymptotic regime.

In this work, we use a simple model to study the change of complexity index $\mu$ (which represents the statistics) of the strongly interacting systems on the level of single and short trajectories. In this way, we propose to identify and isolate those factors that determine how complexity may be regulated. But first, we must develop a new nomenclature to facilitate discussion in this rapidly growing area of study, which is done in Section 2 on *Basic Concepts*.

## 2. Basic Concepts

To make the ingredients of the computation as clear as possible, we review the basic concepts in this new area of network medicine. To generate CE time series it is necessary to note that CEs are a type of renewal event (RE) which is a concept initiated by William Feller[11] wherein he developed the mathematical foundation of RE theory under the assumption that the time intervals between successive discrete events are statistically independent. In addition to being a RE, a CE has an IPL PDF for the waiting-time intervals ($\tau$'s) between events.

Consider an empirical time series $X(t)$ that is a mixture of CEs and non-CEs, for example its being a mixture of crucial and Poisson (or fractional Brownian Motion) events. This is important because it is often difficult to distinguish between the two as both the CEs and the non-CEs are renewal. The typical time series can be written in the general form: $X(t) = \varepsilon X_{\mu<3}(t) + (1-\varepsilon)X_{\mu>3}(t)$. Note that in this empirical relation $0 < \varepsilon < 1$ is the probability that the event at time $t$ is crucial $X_{\mu<3}(t)$ and $(1-\varepsilon)$ is the probability that the event is non-crucial $X_{\mu>3}(t)$. A discrete sequence of CE can be identified using an autocorrelation function which for a genuine CE satisfies $C(t) = 0$ for $t > 0$ and $C(0) = 1$.

The most important new concept is that of *CEs time series* which we have come to understand to be the foundation necessary for the formation of any quantitative theory of medicine or its complimentary theory of rehabilitation, including the formulation of any theory of information transfer among ONs whether one or all are healthy, ill/injured or a combination of the two. Consequently, we pull together the results of studies that have led to this view in Section 2.1. The temporal complexity that is intrinsic to CEs is used to distinguish between CC and CMMEs in Section 2.2 wherein the empirical evidence for many of the conclusions is spelled out.

The final new concept discussed herein is that of emergent intelligence (EI). In Section 2.3, we give an admittedly limited definition of EI and the fact that numerically simulated ON can generate CE time series with even this limited amount of EI we find to be truly remarkable.

## 2.1 Crucial Events
ONs interact within larger networks, which creates a NoONs, and their mutual adaptability is crucial for specific tasks. Such NoONs are coordinating or cooperating, and as such, in order for them to carry out a specific task, each of the interacting ONs should be mutually adaptive to changes in its environment. For example, the environment of an ON consists of the other adaptive ONs within a NoONs and consequently the time series generated by the dynamics of these distinct ONs, e.g., two such ONs within the human body generate the respiration and cardiac datasets, both of which have temporal complexity[23,24], meaning that their statistics are governed by multifractal dimensional (MFD) time series which are generically given by Crucial Event (CE) time series.

CEs are defined by discrete time series with statistically independent time intervals $\tau$ between consecutive events. The key points of CEs have been introduced herein where needed for emphasis and are gathered here for ease of reference:

- CEs are a subset of REs.

- CEs have an inverse power law (IPL) probability density functions (PDFs) for the time intervals between events.

- The IPL PDF is represented as $\psi(\tau) \propto \tau^{-\mu}$, where the IPL index $\mu$ is between 1 and 3. The CETS are ergodic when $3 > \mu > 2$ and non-ergodic when $2 > \mu > 1$.

- The IPL index $\mu$ is a measure of the complexity of the CETS and, as such, is equal to the MFD $\mu(t) = D(t)$.

- A second IPL for CEs time series appears in the Power Spectrum Density (PSD) in terms of the frequency $f$: $S(f) \propto f^{-\beta}$, where the IPL index in this case is $\beta = 3 - \mu$ and ranges between 0 and 2.

- The PSD IPL index at the value $\beta = 1$ gives rise to 1/f-noise at the border between ergodic and non-ergodic time series ($\mu = 2$)[16] whereas for other values of $\mu$ we refer to the CETS as having $1/f-$variability[25].

The utility of CEs has been found in the modeling of neurons firing[26] and in the analysis of network reliability and maintenance, generally modeled with time-dependent parameters. These and many more engineering considerations are presented in the brilliant work of Cox[27] and were applied to medical phenomena by West et al.[21,25]. For most physiological data, the IPL index lies in the interval $2 < \mu < 3$, with $\mu \simeq 2$ in a healthy brain[23]. This region is ergodic but the second moment does not exist (i.e., it is divergent). The region of $1 < \mu < 2$ is non-ergodic, and both the first and second moments do not exist. Consequently, one should be cautious in using measures based on the average or second moment (e.g., detrended fluctuation analysis, DFA) when dealing with complex data in medicine as well as elsewhere.

## 2.2 CC differs from CMMEs
CMMEs[20] identifies the transfer of information between two weakly connected systems as being maximal when the complexity level identified by the MFD time series of two interacting systems is the same[28]. The weak interaction between such systems was considered as a perturbation in a way that the complexity of the drive and the driven systems remain constant. It was studied using the linear response of a system to a time-dependent perturbation using the theory developed by Kubo[29], which has a quantum mechanical origin.

By way of contrast, Mahmoodi and West et al.[17,18] showed that the scaling of simultaneously recorded time series from physiological ONs, e.g., the brain, heart, and lungs, obtained using a modified DEA processing technique produces distinct scaling time series, i.e., varying complexity, in synchrony with one another, which we have designated as CS. This indicates a strong mutual interaction between these systems beyond the limits of linear response theory. In[30], we studied the time series from model phenomena using datasets generated by interacting Selfish Algorithm agents (SA-agents) and showed that CS is a general emergent property of mutually adaptive systems, strongly interacting, whether using empirical physiological datasets or model-generated surrogate data. The CS between SA-agents having the same mathematical infrastructure as that of physiological ONs suggesting the hypothesis that SA-agents could be used as trainers, rehabilitators, or decision-makers, to form successful human-machine hybrid teams[30].

A recent experiment on rehabilitation as a change of complexity was done by Almurad et al.[31] where they postulated and then verified that when an elderly person walks arm-in-arm with a young care-giver the strong interaction provides the mechanism with which to restore their complexity. After three weeks of walking arm-in-arm for an hour a day, they observed a restoration of complexity in the elderly within the walking couples which persisted for at least two weeks beyond the training session. Their work presents the first demonstration of a restoration of complexity in complexity deficient NoONs.

The change in the complexity can be explained by means of the Shannon/Wiener information entropy, i.e., between two strongly interacting networks, the information-poor network increases in (information) complexity until it is on a par with the information-rich network. The exchange of information does not follow the energy gradient and apparently violates the traditional form of the Second Law of Thermodynamics, but this is not the case as explained below.

When two physical networks of approximately the same size are brought into contact with one another, the hotter network loses energy to the cooler network until they reach thermal equilibrium, and the temperature is uniform across the two. Information transfer does not work this way. It is true that the information-rich network increases the complexity level of the information-poor network through the transfer of information, doing so without necessarily losing any of its own complexity, and therefore its information content might remain the same.

### 2.3 Mutual Adaptive Environment Breeds Intelligence

Mahmoodi, West and Grigolini (MWG)[12,13] were apparently the first to use mutually-adaptive agents to create emergent intelligence. Each agent in their models plays a social dilemma game, such as the Prisioner's Dilemma game, with other agents and tries to optimize its payoff, using a simple reinforcement learning mechanism, along with other agents, could resolve the dilemma. The resolution of the dilemma is what is called emergent intelligence that is because of the mutual interaction of adaptive agents. More importantly, the bottom-up emerged intelligence leads back the individual's behavior, that is, whenever an agent wants to free ride and get the advantage of the cooperative environment, the environment, which is composed of other adaptive agents, reacts and forces that agent, through decisions such as not playing or playing as a defector with it, to bounce back to cooperative behavior. This flow of leadership from bottom to top and vice versa keeps the organization robust.

## 3. Model and Methods

To model the strong interaction between mutual adaptive systems, we use surrogate time series with given complexity $\mu$ to represent their dynamics and then define the way they interact through learning and training, which results in change in their complexity. For simplicity, since the present analysis is of the 'proof-of-principle' kind, here we restrict the study to unidirectional interactions.

### 3.1 Manneville Map

To generate a surrogate time series with a given temporal complexity we use idealized Manneville map[21], which produces the time intervals between CEs in a CE time series, denoting the time interval by $\tau$. The waiting time PDF, $\psi(\tau)$, of these $\tau$s have a IPL index $\mu$, or, equivalently, their survival probability, i.e., $\Psi(\tau) = 1 - \int \psi(\tau) d\tau$ can be shown to have the form:

$$\Psi(\tau) = \left(\frac{T}{T+\tau}\right)^{\mu-1}, \tag{1}$$

where $T$ is a small time constant and the IPL index $\mu$ is the complexity measure. Since $\Psi(\tau)$ is a number between 0 and 1, we can set it equal to a random number $k$ between 0 and 1 and find its corresponding $\tau$. Indeed, we can solve equation 1 for $\tau$ as follows:

$$\tau/T = -1 + \frac{1}{r^{\frac{1}{(\mu-1)}}} \tag{2}$$

where $r$ is the random number replacing $\Psi(\tau)$. Consequently, using this map, for any given random number we can get a $\tau$ from the waiting-time distribution of complexity $\mu$.

### 3.2 Learning and Training

One of the few books that directly addresses the statistics of how learning and training are related to the complexity of the thing being learned is *Empirical Paradox, Complexity Thinking and Generating New Kinds of Knowledge*[32] and how that complexity determines the distribution of the mastery of what is or can be learned. No matter how hard you train there is probably a threshold beyond which you cannot go and O'Boyle and Aquinis[33] explain the empirical truth of this observation in the following way[32]:

> O'Boyle and Aquinis conducted five studies involving 198 samples, including over 600,000 entertainers, politicians, researchers, as well as, amateur and professional athletes. Their results are consistent across industries, types of jobs, types of performance measures, and time frames, and indicate that individual performances follow a IPL Pareto, not a normal PDF.

Thus, the empirical results of O'Boyle and Aquinis constrain all theories and applications that address the learning and performance of individuals, including performance measurement and management, training and development, personnel selection, leadership, and the prediction of performance.

The IPL PDF of a CE time series has a finite average value when the IPL index is in the interval $2<\mu<3$ but this cannot be used to characterize the variability in the data. The average value of the complex property being measured, say in a competitive field of human activity, such as sports or science research, usually occurs in the 4th percentile or so. Those that compete for medals in the Olympics, represent the outstanding few; as do those that receive the Noble Prize in one of the sciences after being in the top contenders in a National Academy of Science; etc.

To model the learning and training in a complex system, first we create a surrogate time series with a given complexity index $\mu$ as follows: Using the Manneville map we break a time interval $\tau_1$ to three significant parts; set the first and last part of the trajectory at zero and for the rest of the time interval the trajectory is randomly set to a value of 1 or -1 (see equation 3 and the black trajectory in Figure 2). We continue this process by generating more $\tau$s until the length of the whole time series reaches the value $L$. This time series mimics the dynamical behavior of a complex network measured by an IPL index $\mu$.

Equation 3 shows a section of the trajectory for a given time interval of $\tau_{D,i}$, derived from the Manneville map. The signum symbol $Sgn_{D,i}$ denotes the sign of the trajectory during the period $\tau_{D,i}$, which is chosen randomly to be +1 or -1. The horizontal parts of the trajectory represent the intervals in which the system retains its decision (+1 or -1), and the parts on the origin represent CEs when the system is sensitive and can be affected by its environment to change decisions. In other words, when this trajectory is close to +1 or -1, it is rigid and unable to learn, but it can strongly affect the other system (i.e. train). On the other hand, when the trajectory is close to zero, it is vulnerable and can be under the influence of the other system (i.e. learn).

Note that we can set more of such intermittent states between +1 and -1 for each given $\tau$, but they quantitatively give the same results as of this work, so, for simplicity, we use only one state as the CE for each $\tau$, i.e., its last part.

### 3.3 Interaction

After setting the mechanism for the time evolution of the trajectories, we let one of them (the driven, S) interact with the other (the driver, D):

$$x_D(t) = x_D(t-1) + \begin{cases} Sgn_{D,i} & t=1 \\ 0 & 1 < t \leq \tau_{D,i} - 1 \\ -Sgn_{D,i} & t=\tau_{D,i} \end{cases} \qquad (3)$$

$$x_S(t) = x_S(t-1) + \begin{cases} \alpha + Sgn_{S,j} & t=1 \\ \alpha & 1 < t \leq \tau_{S,j} - 1 \\ \alpha - Sgn_{S,j} & t=\tau_{S,j} \end{cases} \qquad (4)$$

$$\alpha = k\big((1+x_D)(1-x_S) - (1-x_D)(1+x_S)\big) = 2k(x_D - x_S) \qquad (5)$$

We set $x_D(t)$ (with complexity index $\mu_D$) as the driver system, $x_S(t)$ as the driven system (with complexity index $\mu_S$) and define their connection by equations 4 and 5. The quantity $\alpha$ shows how the driven system reinforces its trajectory in order to minimize its distance from the driver's trajectory, where $k$ is the interaction's/learning's coefficient.

The rationale for Equation 5 can be explained by assuming that each system consists of units that can be in one of two possible states, A or B. The change in the population of the driven system, under the influence of the driving system, is proportional to the rate at which its units in state $A_S$ transition to state $B_S$ (i.e., the ratio of units in state $A_S$ multiplied by the ratio of units in state $B_D$), minus the rate at which units already in state $B_S$ revert to state $A_S$ (i.e., the ratio of units in state $B_S$ multiplied by the ratio of units in state $A_D$.) With some simple algebra, this leads to Equation 5.

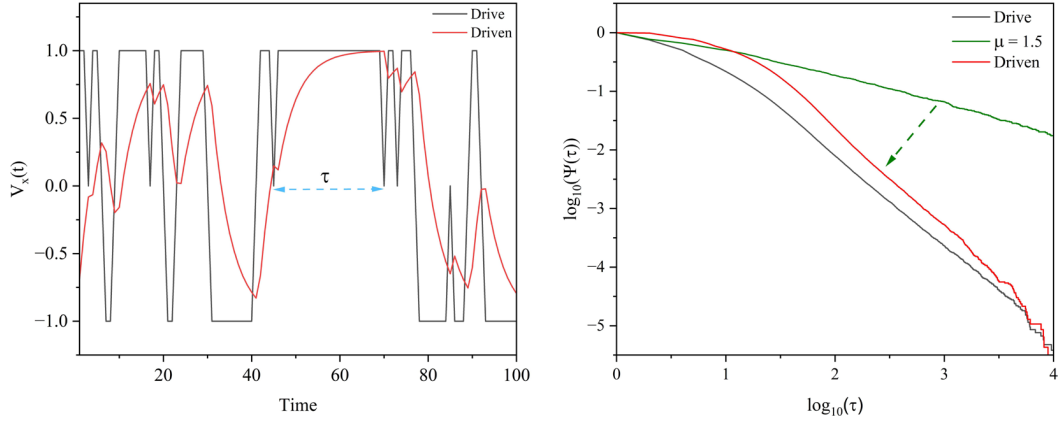

**Figure 2.** Left panel: The black and red curves are, respectively, a section of the time series of the driver and the driven systems. The driven system is connected to the driver system using equations 4 and 5. $\mu_D = 2.5, \mu_S = 1.5$, with interaction coefficient $k = 0.05$. The red curve is the learning curve of the driven system. The dashed blue line shows a time distance $\tau$ generated from Manneville Map. Right panel: The black and red curves are the survival probabilities of the time intervals between two consecutive crossings of the time series of the driver and driven systems. The green curve is the survival probability of the time intervals between two consecutive crossings of an isolated system with $\mu_S = 1.5$.

As a boundary condition, we set $x_S(t)$ to $x_S(t) * sign(x_S(t))$ if $|x_S(t)| > 1$. According to equation 5, the driver system could have high impact on the driven system if its decision is nearly +1 or -1 and at the same time the driven system's decision is nearly 0. In other words, the driven system modifies itself through what is called reinforcement learning in CyberPsychology literature.

To measure the complexity of the driver and driven systems, we evaluate the survival probabilities of the time intervals between the two consecutive crossings of the origin of their time series and the slope of the linear part of the graph of the survival probability, on a log-log graph, determines $\mu - 1$, as shown in the right panel of Figure 2.

To study CS, we set the complexity of the driver system to smoothly changeover time, with periodicity P, as $\mu_D(t) = A + B * cos(t/P)$ and let the driven system, with a given complexity index $\mu_S$, to interact with it. To track the changes in $\mu_S$, we slice the $X_S(t)$ into windows of the length of $3 * 10^5$ (with an overlap of $1 * 10^5$) and evaluate the scaling for each of them as explained above.

## 4. Results

The left panel of Figure 2 shows part of the time series of the driver (black) and the driven (red) systems, where the driven system (with $\mu_S = 1.5$) is connected to the driver system (with $\mu_D = 2.5$) with interaction coefficient $k = 0.05$. The right panel of Figure 2 shows the corresponding survival probability of the time distances between two consecutive events (crossing the origin) of the black and red curves of the left panel, each of them with length $L = 2 * 10^7$. The slopes of the linear sections of the curves measure -1. The blue curve is the survival probability for an isolated system with $\mu_S = 1.5$. Comparing the blue curve with the red curve, which is the survival probability of the driven system with initial IPL index $\mu_S = 1.5$, shows the complexity change because of the interaction.

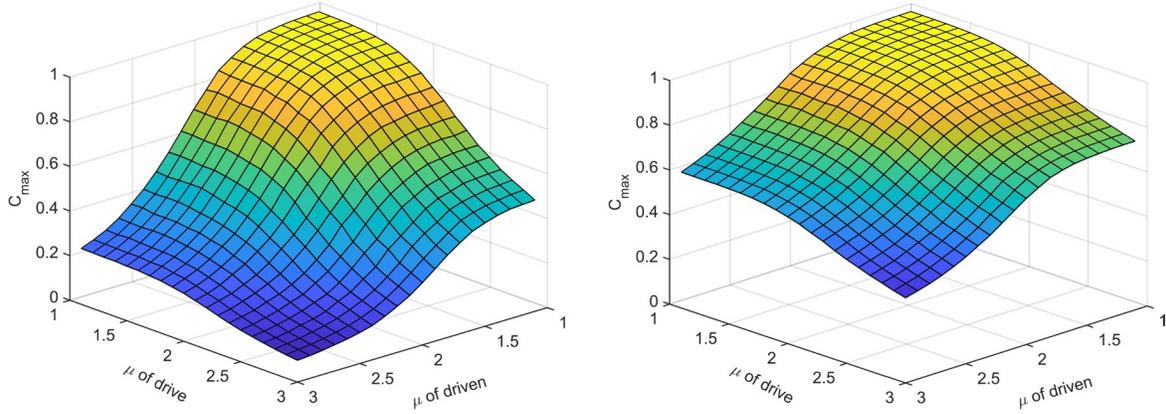

**Figure 3.** The maximum of the cross-correlation between the time series of the driver and driven systems with different values of $\mu_D$ and $\mu_S$. In the left and right panels, the interaction coefficient $k$ is 0.01 and 0.05, respectively. See text for an interpretation of the contrasts in the paneled results. Ensemble = 1.

Figure 3 shows the maximum of the cross-correlation between the driver and driven time series with different complexity indexes $\mu$. The driven system is connected to the driver system with an interaction coefficient of $k = 0.01$ (left panel) and $r = 0.05$ (right panel). We did not use any ensemble averages to evaluate the cross-correlations.

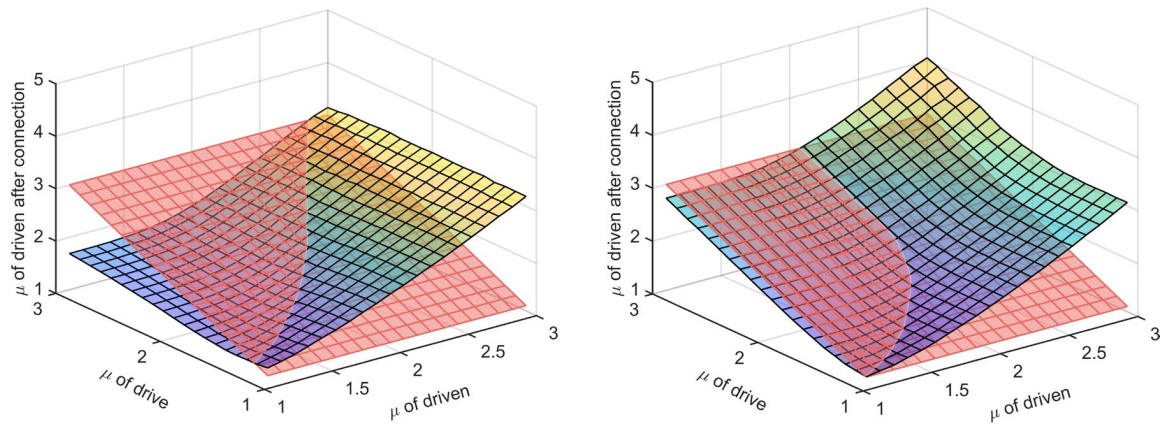

**Figure 4.** The complexity as measured by the index $\mu$ of the driven system after connection to the driver system with complexity index $\mu_S$ and $\mu_D$, respectively. In the left and right panels, $k$ is 0.01 and 0.05, respectively. The light red plane shows the $\mu$ of the driven system if the changed to that of the driver system. Ensemble = 1.

Figure 4 depicts the change in the complexity of the driven system after interacting with the driver system for weak (left panel, $k = 0.01$) and strong (right panel, $r = 0.05$) interaction. The light red plane shows the complexity index of the driven if its complexity changed to that of the drive. While the weak interaction could not change the complexity of the driven, the higher interaction curved the complexity plane of the driven systems, especially for $\mu_S < 2$, to reach the complexity of the drive.

The panels of Figure 5 show the CS between the driver system with a changing complexity $\mu_D(t) = A + B \ast cos(t/P)$ (blue curve), and the driven system (red curve) with complexity $\mu_S$ for different values of A, B, and $\mu_S$. On the left we have two dyadic interacting complex networks with the blue curve in the upper left panel has the driver $\mu_D < 1.5$ in a small interval straddling the negative peak of the cosine and contrast this with the more regular CS in the lower left panel where the driver satisfies $\mu_D > \mu_S = 1.5$ throughout the cycle. The curves depicted in the upper right panel increases the response complexity index to $\mu_S = 2.5$ such that $\mu_D < \mu_S$ throughout the divers' cycle. The lower right panel has the diver shifted upward by 0.5 in the initial value of $\mu_D = 2.75$ at $t = 0$. The CS effect is still maintained but having the driven system with a $\mu_D < \mu_S$ for even a small fraction of the drivers' period distorts the CS making it appear noisy but still recognizable.

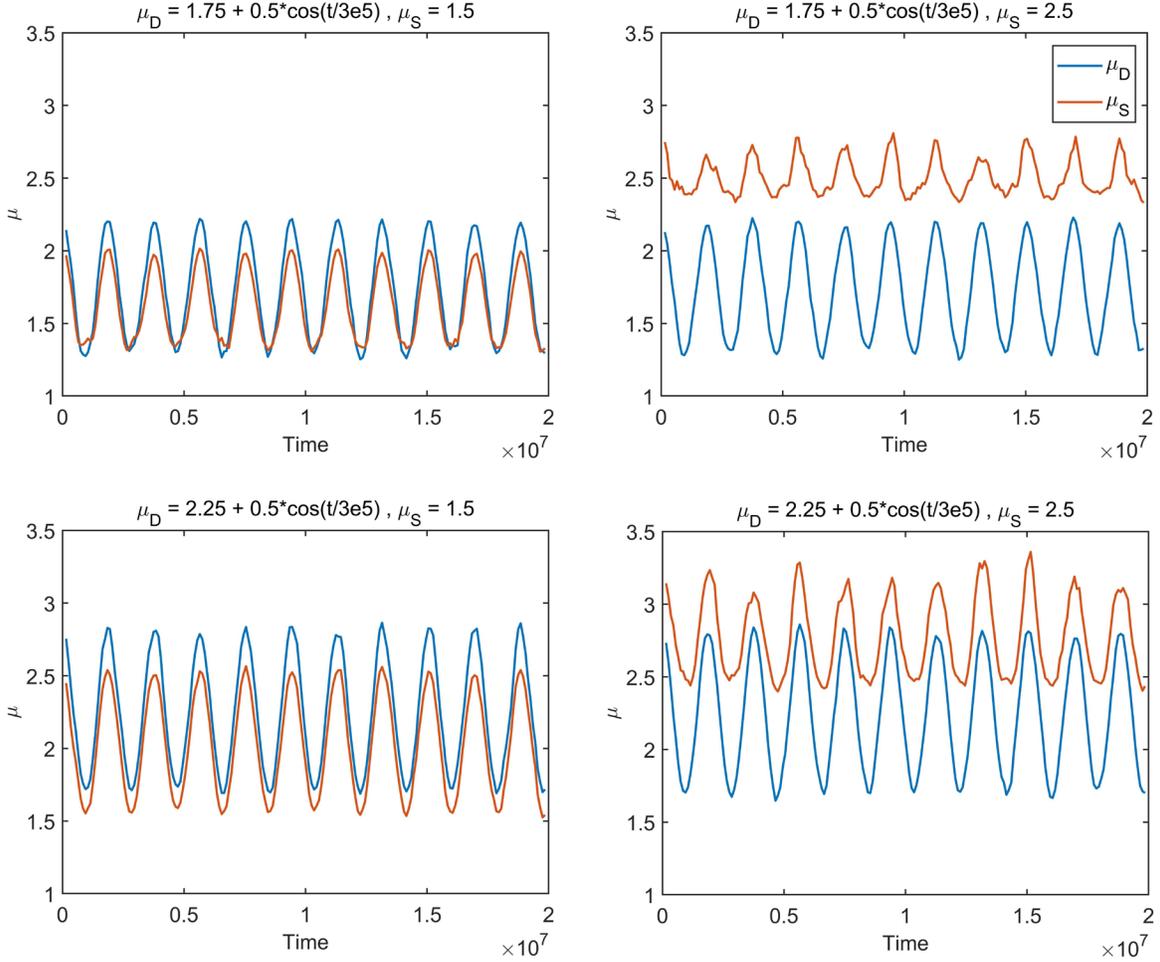

**Figure 5.** Each red curve shows the complexity index $\mu$ of the driven system after being connected to the driver system with a changing complexity index $\mu_D(t) = A + B\cos(t/P)$. B= 0.5 and P = 3 ∗ 10$^5$. In the top and bottom panels, A = 1.75 and 2.25, respectively. In the left and right panels $\mu_S$ is 1.5 and 2.5, respectively. The cross-correlations between the red and blue curves are 0.9896, 0.9195, 0.9979, and 0.9721 for the top-left, top-right, bottom-left, and bottom-right panels, respectively. Ensemble = 1.

## 5. Discussion

In this section we provide comments to bring insight into the results depicted in Figures 3 -5 that can facilitate our understanding of the change in the complexity as has been shown experimentally in the walk arm-in-arm[34] and in empirical time series of EEG, ECG, and Respiratory datasets using newly developed data processing techniques[17,18], as well as in theoretical calculations using an agent-based model[35].

We first created a surrogate time series with a given complexity index $\mu_D$ as the driver system. Such time series mimic the dynamics of intelligent groups. Then we connected this intelligent (educated) driver to another surrogate time series a driven system having a lower complexity index. This second system has the capability of learning and adapting itself to the driver system and is therefore the (uneducated) driven system. This interaction allows us to study the changes in the relative complexity of the two strongly interacting systems.

Figure 3 shows that the driven system can follow the pattern of the driver the best if both of them have a complexity index in the range of 1 to 2, i.e., both of them are non-ergodic (yellow plateau). The second most influential cases are when one of them is non-ergodic, and the other is ergodic (orange and green bands). The least effective configuration is when both of them are in the ergodic regime (blue plateau). Also, this figure reveals that increasing the intensity of the interaction strength increases the ability of the driven to follow the driver. The comparison of Figure 3 with Figure 1 shows that CC and CMMEs are distinct phenomena. We stress that for the results of the present work, we do not use any ensemble averages (neither in time nor in ensemble number), which is essential for the determination of the CMME.

Panels of Figure 4 show that to change the complexity of the driven system, the intensity of the interaction must be sufficiently large. Increasing the interaction coefficient $k$ forces the complexity plane curve to rest on the plane of complete matching as shown by the growth in the intersection of the two planes (red and blue) in the left panel into an overlap region of the two planes in the right panel. These results can explain how rehabilitation occurs in the arm-in-arm experiments of Almurado et al.[34].

Figure 5 shows that if the complexity of the driver system changes over time, as experimentally shown in the complexity index of the EEG data by Mahmoodi, West et al.[17,18] or in agent-based models by us[30], and such changes can be projected onto the complexity plane of the adaptive driven system. This effect can explain the CS effect and shows how the CS depends on the relative distance between the complexity indices of the two systems.

Note that we[17,18,30] used Modified Diffusion Entropy Analysis (MDEA) to evaluate the scaling index $\delta$, which is related to the temporal complexity measure $\mu$ by equations $\mu = 1 + \delta$ for $1 < \mu < 2$ and $\mu = 1 + 1/\delta$ for $2 \leq \mu < 3$, respectively. Herein we directly evaluated $\mu$ from the survival probability of the time distances between the crossings of the time series (i.e., $\tau$'s), since we used sufficiently long time series to provide us with enough $\tau$'s to make good statistics.

The driven system in our model is a simple agent that adapts itself to the driver system, with a complexity comparable to that of physiological systems, by reinforcement learning. This suggests that by using such adaptive agents we can design machines to enhance the intelligence of physiological systems or social groups as rehabilitation/training, or, on the contrary, we can lower the intelligence of such systems if they are adversaries, such as cancer cells or foes.

## Conclusion

In Mahmoodi and West et al.[17,18] we showed that CS exists between ONs, and subsequently, in Mahmoodi et al.[30] we showed that CS is an emergent property of systems composed of mutual interacting components. In the present sequel to those previous works, we presented herein a simple model based on CE time series and reinforcement learning that led to a deeper understanding of CC as well as providing additional insight to better explain the CS effect.

The potential applications of the concept of CC are, for example, in a proper way for training, teaming, and rehabilitation. In general, the difference between CMMEs and CC can lead to the designing of adaptive non-invasive/invasive signals to increase/decrease the complexity (intelligence) of the targeted sub-network in order to improve the performance of the whole system.

In spite of the three-quarters of a century of research into *cybernetics*, which was Norbert Wiener's vision of how to optimize the man-machine interface, little is known about how to observe, manage, and improve biological/artificial/ hybrid human-machine interactions. Due to its relative lack of progress, this area of research might, with a change of focus, still be considered new and be called '*cybernetic − conscious − communication*'. This marriage of information technology (IT) with emergent intelligence within AI-agent-based networks to mimic consciousness, if successful, would revolutionize how humans work with computers and how computers work with humans.

## Code availability

All the codes used to produce the results of this work are available at https://github.com/Korosh137/Complexity-Control.git

## Acknowledgments

Research was sponsored by the Army Research Laboratory and was accomplished under Cooperative Agreement Number W911NF-23-2-0162. The views and conclusions contained in this document are those of the authors and should not be interpreted as representing the official policies; either expressed or implied, of the Army Research Laboratory or the U.S. Government. The U.S. Government is authorized to reproduce and distribute reprints for Government purposes notwithstanding any copyright notation herein.

## Author contributions

K.M conceived CC and conducted the analysis. K.M and B.J.W wrote the draft of the manuscript. All authors critically assessed and discussed the results and revised and approved the manuscript.

## Competing interests

The authors declare no competing interests.

# References

1. Wiener, N. *Cybernetics or Control and Communication in the Animal and the Machine* (MIT press, 2019).
2. Shannon, C. E. & Weaver, W. The mathematical theory of communication, 117 pp. *Urbana: Univ. Ill. Press.* (1949).
3. Von Neumann, J. & Morgenstern, O. *Theory of games and economic behavior: by J. Von Neumann and O. Morgenstern* (Princeton university press, 1953).
4. Ashby, W. R. An introduction to cybernetics. (1956).
5. West, B. J., Turalska, M. & Grigolini, P. *Networks of echoes: imitation, innovation and invisible leaders* (Springer Science & Business Media, 2014).
6. West, B. J. A mathematics for medicine: the network effect. *frontiers Physiol.* **5**, 456 (2014).
7. Zipf, G. K. Human behavior and the principle of least effort: An introduction to human eoclogy. (1949).
8. Auerbach, F. Das gesetz der bevölkerungskonzentration [the law of population concentration]. *Petermanns Geogr. Mitteilungen* **59**, 74–76 (1913).
9. Pareto, V. *Cours d'économie politique*, vol. 1 (Librairie Droz, 1964).
10. West, B. J. & Grigolini, P. *Complex webs: anticipating the improbable* (Cambridge University Press, 2010).
11. Feller, W. *Introduction to Probability Theory and its Applications, 2 Volumes* (Wiley and Sons, NY, NY, 1950).
12. Mahmoodi, K., West, B. J. & Grigolini, P. Self-organizing complex networks: individual versus global rules. *Front. physiology* **8**, 478 (2017).
13. Mahmoodi, K., West, B. J. & Grigolini, P. Self-organized temporal criticality: bottom-up resilience versus top-down vulnerability. *Complexity* **2018**, 1–10 (2018).
14. Mega, M. S. *et al.* Power-law time distribution of large earthquakes. *Phys. Rev. Lett.* **90**, 188501 (2003).
15. Allegrini, P., Grigolini, P., Hamilton, P., Palatella, P. & Raffaelli, G. Memory beyond memory in heart beating, a sign of a healthy physiological condition. *Phys. Rev. E* **65**, 041926 (2002).
16. Grigolini, P., Aquino, G., Bologna, M., Lukovic´, M. & West, B. J. A theory of 1/f noise in human cognition. *Phys. A: Stat. Mech. its Appl.* **388**, 4192–4204 (2009).
17. Mahmoodi, K., Kerick, S. E., Grigolini, P., Franaszczuk, P. J. & West, B. J. Complexity synchronization: a measure of interaction between the brain, heart and lungs. *Sci. Reports* **13**, 11433 (2023).
18. West, B. J., Grigolini, P., Kerick, S. E., Franaszczuk, P. J. & Mahmoodi, K. Complexity synchronization of organ networks. *Entropy* **25**, 1393 (2023).
19. West, B. J., Geneston, E. L. & Grigolini, P. Maximizing information exchange between complex networks. *Phys. Reports* **468**, 1–99 (2008).
20. Aquino, G., Bologna, M., West, B. J. & Grigolini, P. Transmission of information between complex systems: 1/f resonance. *Phys. Rev. E: Stat. Nonlinear, Soft Matter Phys.* **83**, 051130 (2011).
21. West, B. J. & Grigolini, P. *Crucial events: why are catastrophes never expected?* (World Scientific, 2021).
22. Mahmoodi, K., West, B. J. & Gonzalez, C. Selfish algorithm and emergence of collective intelligence. *J. Complex Networks* **8**, cnaa019 (2020).
23. Allegrini, P. *et al.* Spontaneous brain activity as a source of ideal 1/f noise. *Phys. Rev. E* **80**, 061914 (2009).
24. Allegrini, P., Barbi, M., Grigolini, P. & West, B. J. Dynamical model for dna sequences. *Phys. Rev. E* **52**, 5281 (1995).
25. West, B. J., Grigolini, P. & Bologna, M. *Crucial Event Rehabilitation Therapy: Multifractal Medicine* (Springer Nature, 2023).
26. Goris, R., Movshon, J. & Simoncelli, E. Partitioning neuronal variability. *Nat. Neurosci* **17**, 858–865 (2014).
27. Cox, D. *Renewal theory* (Science paperbacks and Methuen and Co. Ltd., 1967).
28. West, B., Geneston, E. & Grigolini, P. Maximum information exchange between complex networks. *Phys. Reports* **468**, 1–99 (2008).
29. Kubo, R. Statistical-mechanical theory of irreversible processes. i. general theory and simple applications to magnetic and conduction problems. *J. physical society Jpn.* **12**, 570–586 (1957).
30. Mahmoodi, K. *et al.* Complexity synchronization in emergent intelligence. *Sci. Reports* **14**, 6758 (2024).
31. Almurad, Z. M., Roume, C. & Delignières, D. Complexity matching in side-by-side walking. *Hum. Mov. Sci.* **54**, 125–136

(2017).

32. West, B. J., Mahmoodi, K. & Grigolini, P. *Empirical paradox, complexity thinking and generating new kinds of knowledge* (Cambridge Scholars Publishing, 2019).

33. O'Boyle Jr, E. & Aguinis, H. The best and the rest: Revisiting the norm of normality of individual performance. *Pers. Psychol.* **65**, 79–119 (2012).

34. Almurad, Z., Roume, C., Blain, H. & Delignieres, D. Complexity matching: Restoroing the complexity of locomotion in older people therough arm-in-arm walking. *Front. Physiol. Fractal Physiol.* **9**, 1–10 (2018).

35. Mahmoodi, K. & Gonzalez, C. Emergence of collective cooperation and networks from selfish-trust and selfish-connections. In *CogSci*, 2254–2260 (2019).